\documentclass[a4paper]{jpconf}
\usepackage{graphicx}
\usepackage{cite}
\begin{document}
\title{Overcoming High Energy Backgrounds at Pulsed Spallation Sources}

\author{ 
Nataliia Cherkashyna$^1$,
Richard J Hall-Wilton$^{1,2}$,
Douglas D DiJulio$^1$,
Anton Khaplanov$^1$,
Dorothea Pfeiffer$^{1,3}$,
Julius Scherzinger$^{1,4}$,
Carsten P Cooper-Jensen$^{1,5}$,
Kevin G Fissum$^{1,4}$,
Stuart Ansell$^{6}$,
Erik B Iverson$^7$,
Georg Ehlers$^8$,
Franz X Gallmeier$^8$,
Tobias Panzner$^9$, 
Emmanouela Rantsiou$^9$,
Kalliopi Kanaki$^1$,
Uwe Filges$^9$,
Thomas Kittelmann$^1$,
Maddi Extegarai$^1$,
Valentina Santoro$^1$,
Oliver Kirstein$^{1,10}$ and
Phillip M Bentley$^{1,5}$
\\
}

\address{1 European Spallation Source ESS AB, SE-221 00 Lund, Sweden\\
2 Mid-Sweden University, SE-851 70 Sundsvall, Sweden\\
3 CERN, CH-1211 Geneva 23, Switzerland\\
4 Division of Nuclear Physics, Lund University, SE-221 00 Lund, Sweden\\
5 Department of Physics and Astronomy, Uppsala University, 751 05 Uppsala, Sweden\\
6 ISIS, Rutherford Appleton Laboratories, Harwell Oxford, Didcot  OX11 0Q, United Kingdom\\
7 Instrument and Source Division, ORNL, Oak Ridge, TN-37831, USA\\
8 Quantum Condensed Matter Division, ORNL, Oak Ridge TN-37831, USA\\
9 Paul Scherrer Institute, 5232 Villigen PSI, Switzerland\\
10 The University of Newcastle, Callaghan NSW 2308, Australia}

\ead{nataliia.cherkashyna@esss.se}

\begin{abstract}
Instrument backgrounds at neutron scattering facilities directly affect the quality and the efficiency of the scientific measurements that users perform.  Part of the background at pulsed spallation neutron sources is caused by, and time-correlated with, the emission of high energy particles when the proton beam strikes the spallation target. This prompt pulse ultimately produces a signal, which can be highly problematic for a subset of instruments and measurements due to the time-correlated properties, and different to that from reactor sources. Measurements of this background have been made at both SNS (ORNL, Oak Ridge, TN, USA) and SINQ (PSI, Villigen, Switzerland). The background levels were generally found to be low compared to natural background. However, very low intensities of high-energy particles have been found to be detrimental to instrument performance in some conditions. Given that instrument performance is typically characterised by S/N, improvements in backgrounds can both improve instrument performance whilst at the same time delivering significant cost savings. A systematic holistic approach is suggested in this contribution to increase the effectiveness of this. Instrument performance should subsequently benefit. 
\end{abstract}

\section{Introduction}

In this contribution during ICANS XXI, the results of both simulation studies and experimental surveys trying to verify shielding at a basic level were shown, relating to the consideration of future instrument shielding at European Spallation Source (ESS) which is currently under construction in Lund, Sweden~\cite{TDR}.
In the spirit of ICANS, this writeup will not concentrate on the state of detailed simulations or surveys themselves, as these are better written up in individual journal contributions when complete, but rather try and give a perspective on how these important studies might fit together, and to give a few philosophical musings on what might be the most effective way to do it.
It should be noted that these musings should not be taken as unique, but rather fit into a field of approaches, some of which were also shown at ICANS~\cite{icansxxi-shielding}. Shielding is very much topical!

In contrast, neutron optics and transport has made great strides during the 90's and 00's --- advances in neutron optics ~\cite{neutron-guides} being instrumental in this. 
Very roughly speaking this is about transporting maximal brilliance transfer of the desired phase space of neutrons to the sample position. 
Coupled with this development, and with the coincident rise~\cite{moore} and democratisation~\cite{lyons} of computing power, are the now ubiquitous~\cite{weiser} and user-friendly ray-tracing programs such as McStas~\cite{mcstas1,mcstas2,mcstasw} and Vitess~\cite{vitess1,vitess2,vitessw} which use Monte Carlo techniques~\cite{montecarlo1} to simulate the thermal and cold neutron transport to have a huge effect on instrument design. 
These techniques have become so ingrained, that no proposed instrument design is believed without having been {\it ``verified''} using one of these simulations to evaluate ideal performance. 

Simply put, these programs are idealistic calculations to optimise phase space transport of neutrons from source to sample and beyond; this is effectively optimisation of the signal at the sample.
However, most existing instruments performance is better characterised by a figure of merit related to some function of signal-to-noise ($S/N$) performance.  
Often the limiting factor at the floor of the noise is the instrumental background.
As such, background calculations are now an essential part in the optimisation process, that needs to be done in harmony with the cold/thermal neutron simulation programs. As will be seen below, it is also not given that signal optimisation is the same as signal-to-background (S/B) optimisation.

\section{Shielding of Backgrounds \label{sec:shielding}}

Shielding is often seen as a {\it ``necessary evil''} for the budget of the instrument; needed for radiological protection purposes, licensing and safety. 
For spallation sources it adds up to a very significant fraction of the instrument budget~\cite{LOKIp,LOKIi, NMX}.
Indeed, more can be spent on shielding than on neutron optics, which is intriguing given how much Monte-Carlo time is often dedicated to the optimisation of the guide shape in preference to the optimisation of the shielding.  This is especially true when adding up the effort and cost of fixing poor experimental backgrounds during instrument operational phases over the lifetime of the facility.
Considering shielding only in the safety context neglects the most important role that shielding has to play in an instrument's performance: maximising the $S/B$ ratio.

Shielding is also typically considered locally at a static point - i.e. in this position what amount of shielding is needed for the present particle fluences.
This neglects the synergies from a holistic approach to its design for an instrument: \emph{i.e.} rather than taking the radiation field as given and absolute for an particular instrument design and post-mortem design the shielding around it; instead the goal should be to try to engineer this radiation field to make it minimal at the points which are key for instrument performance --- normally the detectors and sample position --- and mould the radiation field to allow the quantity ($\propto$ cost) of shielding to be globally reduced.

In terms of materials utilised, only four are typically considered in neutron scattering: 
\begin{enumerate}
\item Concrete. This is taken as a {\it ``light hydrogen containing''} cheap material, primarily to bulk to shield gammas and to moderate fast neutrons. Whilst it is just about the cheapest material~\footnote{Except water and earth which are much under-utilised.}, surprisingly it is not ideal at either of the other two tasks, and the hydrogen content of concrete is not so great.  Impurities in the concrete also activate readily if the concrete is allowed to be inadvertently illuminated by strong neutron beams.
\item Steel. This is taken as a {\it ``heavy, dense''} cheap material to slow and stop fast and high energy neutrons~\cite{muhrer-icansxxi}.  Stainless steels activate to the point of being a hazard, whereas mild steels exhibit regions of neutron transparency in the resonant energy range of 100s keV to around 1 MeV.
\item Plastic.  This is a light material that is used to slow fast neutrons ($<20$ MeV).
\item A boron-containing substance. This is to {\it ``mop up''} thermal and cold neutrons. Typical applications are plates of boron carbide, or plastics made with boric acid that combine absorbing and moderating properties.
\end{enumerate}

Perception of cost and a fear of activation issues leads to the exclusion of many other attractive materials being considered.

Lastly, shielding is a multidisciplinary problem: a variety of solutions have been found for several applications. 
It is therefore  apt to ask whether any parts of the solutions found in other disciplines have a lesson for neutron scattering?

\section{Background Sources}

\begin{figure}[h]
\begin{center}
\includegraphics[width=8 cm]{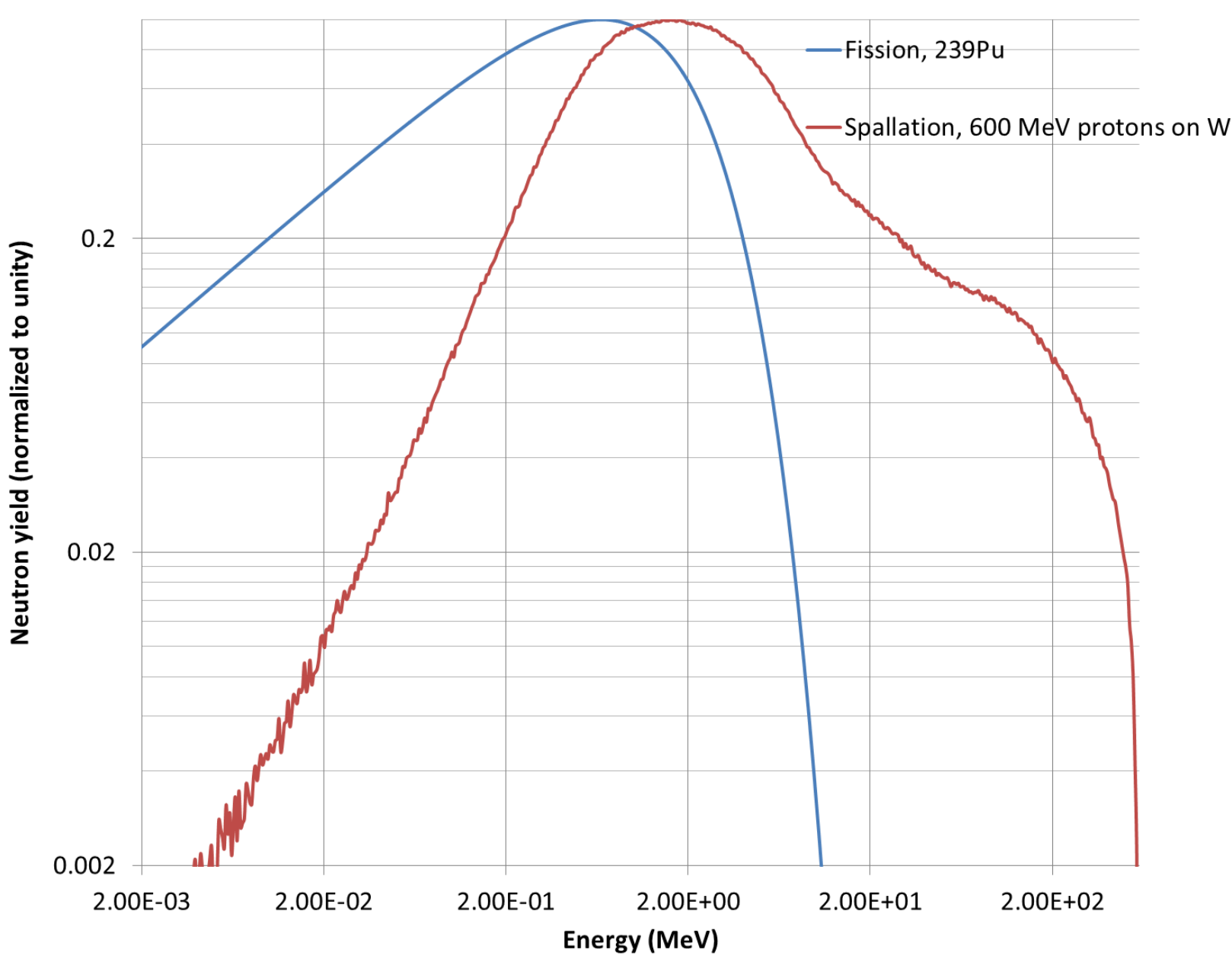}
\end{center}
\caption{Neutron emission spectrum from fission compared to spallation, after Shetty~\cite{shetty_thesis}.}
\label{fig: spallationVSfission}
\end{figure}

One of the most striking pictures illustrating the differences in backgrounds between reactor and spallation sources can be seen from comparing the frequently shown neutron source energy distributions. 
This is shown in figure \ref{fig: spallationVSfission}. 
The difference is significant, with a much longer tail from spallation sources towards higher energies; almost up to the incident beam energy of the proton. 
However, how does this make a difference to the background level?

It should be noted that of course spallation sources are not new: therefore measurements of background at existing sources are important to understand these effects - and just as importantly how they propagate down to instruments - these cascades are important as it is the background at the instrument that is most important. 

To understand these cascades, it is important to review and understand what can happen to these background particles. 

\begin{figure}[h]
\begin{center}
\includegraphics[width=8 cm]{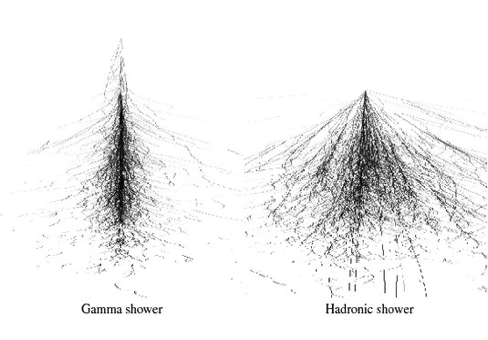}
\end{center}
\caption{The different character of electromagnetic(gamma) and hadronic showers~\cite{gaudio}.}
\label{fig: showers}
\end{figure}

A prime candidate mechanism in the prompt pulse is the phenomenon of particle showers. These are the cascades of secondary particles produced by high energy particle interactions with dense matter. There are two different mechanisms of particle showers: electromagnetic and hadronic~\cite{gaudio}. As shown in figure \ref{fig: showers}, electromagnetic shower is more localised in a material, while hadronic shower is more spatially extended. Electromagnetic showers are caused by photons and electrons. 
Electrons create many events in matter by ionisation and bremsstrahlung, and photons are able to penetrate quite far through material before they lose energy. 
The second type, hadronic showers, is caused by hadronic
particles (baryons and mesons: protons, neutrons, pions, kaons, etc) that are made of quarks - and strong nuclear forces are involved into those interactions. The hadronic showers are characterised by ionisation and interactions between incident particles and nuclei of the material. Hadronic showers are complex, as many reactions take place and many different particles are produced, and it is essential to use sophisticated modelling packages such as Geant4~\cite{agostinelli} to understand hadronic showers.

Some of the interactions of note can be summarised as follows: 
\begin{enumerate}
\item \label{item:photonuclear} High energy gamma photons ($\sim 10-50$ MeV) readily liberate neutrons via GDR photonuclear processes~\cite{gdr-photonuclear}. 
\item \label{item:primaryHadrons} Inelastic scattering of high energy (100s MeV and GeV) neutrons in shielding produces secondary hadrons (mostly neutrons).
\item \label{item:fastNeutrons} Fast neutrons are readily scattered through gaps, light materials, down guide tubes and pipework.  The scattering ``reflectivity'' of most materials to fast neutrons is between 40 and 60 \%.
\item \label{item:meanFreePath} The mean free path, and the tenth value (amount of shielding to reduce the fast neutron flux by a factor of ten) can be quite large at spallation source energies.  Typically, 1 metre of concrete is needed to attenuate a GeV neutron flux by a factor of 10.  Where reactors may be able to shield with a few tens of cm of material, at higher energies these thicknesses can prove to be insufficient with dramatic results, since these in turn can be readily reflected (point \ref{item:fastNeutrons}), generate secondary neutrons 10s--100s of metres from the escape point and potentially inside other instruments (point \ref{item:primaryHadrons}) after they have escaped.
\end{enumerate}

Point \ref{item:photonuclear}, and particularly point \ref{item:primaryHadrons} above, generate showers that produce neutrons.  The primary and secondary neutrons can travel large distances, and lose energy in the instrument shielding, generating more secondaries in the process.  Because the count rates can be quite low, and because the primaries are of such high energy compared to thermal neutron energies, it is not straightforward to measure the primary radiation intensity with most thermal neutron detectors, making it difficult to identify the primary source.  Once the neutrons lose enough energy into the keV -- 1 MeV region, the resonances in steel make steel shielding far less effective.  Where the primaries are escaping seems to vary from facility to facility, so it is important to consider potential accelerator, accelerator-to-target, target, beamline sources.  Moreover, the beamline contributions can be from optics shimming/collimation/ ``horse-collars'', chopper pits, shielding stacking, conduits, and beamstops.

Point \ref{item:fastNeutrons} above means that care has to be taken in evaluating gaps in the shielding, and how the beamline is laid out.  Ensuring that primary energetic neutrons escape into bulk shielding, both around the beamlines and in the beamstops, whilst preventing secondary propagation to the instruments is critical.  This can be in the form of ``get lost'' tube concepts for beamstops, and carefully designed pockets of empty space or low density material around the neutron guides between the heavy collimator blocks.

Hence, {\it effective `line of sight'} differs for fast and thermal neutrons and for gammas. 
The transport of all three species varies significant for different materials.  
Even if line of sight is blocked to the source for visible photons, fast neutrons can scatter through the neutron guide glass and vacuum system easily, and can propagate further down the curved guide, particularly if secondary neutrons are emitted.  It is important to over-engineer the line of sight criterion.
Therefore {\it `line of sight'} becomes a much more {\it mushy} and malleable criterion than a simple geometrical definition.


\section{Background Simulation}

The ray-tracing programs McStas~\cite{mcstas1,mcstas2,mcstasw} and VITESS~\cite{vitess1,vitess2,vitessw} are frequently-used Monte Carlo~\cite{montecarlo2} techniques to simulate the thermal and cold neutron transport. 
They do not however also simulate instrument backgrounds from fast neutrons or gammas. 
They must therefore be combined with additional simulations dedicated to modelling these backgrounds. 

The steps needed to optimise shielding are as follows:
\begin{enumerate}
\item Simulation of phase space of interest - i.e. McStas/VITESS simulation of thermal/cold neutrons. This simulation can partially simulate some background effects, such as multiply-scattered neutrons or neutrons scattered in ancillary equipment near the sample or detectors.
\item Background simulation using specialised high energy/nuclear tools.  
\item Normalise these appropriately and sum these to give the $S/B$ for the instrument --- typically the best estimator for a figure of merit for instrument performance. 
\item Ensure that the critical parts of this simulation are validated. In particular ensure that the simulation results themselves are both sensible (back of envelope) and cross validated with a second method.
\end{enumerate}

Background simulations might involve MCNPX, which is a general-purpose Monte Carlo code that can be used for neutron, photon, electron, or coupled neutron/photon/electron transport~\cite{mcnpx,mcnpxw}.  Equally it could use Geant4, which is a C++ toolkit for creating simulations of the passage of particles through matter and electromagnetic fields, and has areas of application such as high energy, nuclear and accelerator physics~\cite{agostinelli}, including the detector response to this background.  Another alternative could be PHITS --- a general purpose Monte Carlo particle transport code written in FORTRAN. This code deals with the transport of all types if particles within the wide energy range up to 100s of GeV. Several nuclear reaction models and nuclear data libraries are available~\cite{phits}. 

Another simulation tool is CombLayer~\cite{comblayer}, which provides a means of rapidly producing complex MCNPX geometries, that depend on a long list of ranged variables, whilst at the same time optimising the MCNPX input to run quickly. It is also intended to help with placement of tallies, maintaining consistent materials and some variance reduction.

Geant4 is used to perform the simulations within the frame of instrument background studies and detector physics at ESS. That toolkit is a well-known main simulation engine for high energy physics applications. Different materials were investigated in order to understand their shielding properties against high energy particles. The toolkit includes different physics models and processes (electromagnetic, hadronic, optical), tracking, hits, large set of geometrical features, wide energy range.

The physics list recommended for shielding applications  in the energy range of interest is QGSP BERT HP, it is chosen for the shielding studies performed recently. This is the set of physical processes which describe the interaction of particles with dense matter~\cite{agostinelli}. The abbreviation means the following: QGS is for quark-gluon string model, P means precompound, BERT means that Bertini intra-nuclear cascade model is used~\cite{heikkinen}, and HP states that high precision neutron tracking model is also included~\cite{speckmayer}.

For ESS, to ensure that such studies are done in a consistent and rigorous fashion, using known processes, a framework based around Geant4 has been setup precisely for these shielding, detector and instrument design studies~\cite{TK-CHEP,TK-IEEE}. Of particular note is that this includes additions to Geant4 to ensure that the processes relevant to thermal neutron scattering are implemented and much better described by the simulation~\cite{NXSG4,XX-IEEE}.

As part of the validation process, the best strategy is to measure and understand observed backgrounds at extant running spallation sources.
Here it is important to have an open honest dialogue, to be able to learn lessons from existing sources. 
An extensive measurement program has been carried out; as the results this far show it is impressive how good the shielding is in general.

The radiation protection adage {\it ``distance, shielding, time''} has applications also to background reduction. 
The following 2 sections deal with aspects on the ``distance'' and ``shielding'' parts of this adage. 
Utilising the third, ``time'', is viable for a reactor source, however for a pulsed spallation source, it is difficult to utilise this for background minimisation, when the source pulse defines so much about the temporal features in the background.
Combined with time-of-flight and the distances on the instrument, it defines the background time characteristics. 
This often leads to a prompt pulse effect, making sensitive measurements not possible during this time period, or requiring that measurements use flight times less than the repetition period of the source. 
Such deadtimes are less tenable for a long pulse spallation source, such as ESS, necessitating a greater reduction in background levels.


\section{Geometrical Effects}

Simple geometrical effects on the radiation fields may be highly important. 
It is important to remember how these scale: 
\begin{itemize}
\item Point sources, such as the sample, beam stop, or a deformity. The radiation field scales as 1/r$^2$ with distance.  
\item Beamlines. The radiation field scales as 1/r with distance.  
\item Surfaces, such as the monolith wall or loss along a large area of guide wall. To first order the radiation field does not scale with distance; it scales slowly based upon boundary conditions of the surface illuminated and attenuation. 
\end{itemize}
Understanding and engineering such geometries can be used to either reduce the amount of material needed, or to concentrate losses in particular locations to reduce shielding needs elsewhere. 

Another type of structure is the {\it ``get-lost tube''} as mentioned previously, which transports the unused neutrons after the sample away from the detectors, before dealing with them in an appropriately designed beamstop system. 
This has proven to be highly useful for background reduction of neutrons backscattered around the detectors, after passing through the sample. 

There is a prescription that the sample should be out of line of sight. 
But as already formulated earlier: {\it What does line of sight mean? What is the effective line of sight?}
Even more importantly - is line of sight the same for different particle species? 

This concept can be adapted also to try and reduce background from line of sight losses; by optically removing the signal neutrons from the line of sight, and letting high energy gammas and fast neutrons go straight into a controlled loss area. 
By geometrically directing these background particles to known loss points, scatter and albedo from these losses downstream towards the sample position can be vastly reduced. 
In many ways such a geometry for dealing with different energy and particle species is very much analogous to a synchrotron, where one particle type (eg electrons) is bent around the ring, whilst the light produced in this curve is produced in a straight line form the bending point, and off-momentum electrons take a different (typically tighter) orbit, and so can be cleaned by specific collimination on the inner side of the ring. 
All three particle types take a different physical flight path. 

\begin{figure}
\begin{minipage}[t]{19pc}
\includegraphics[width=19pc]{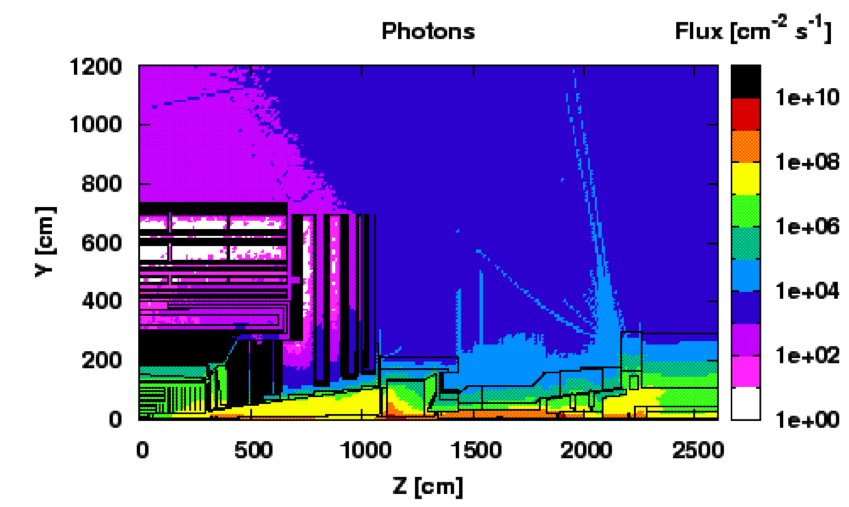}
\caption{\label{fig:CMSlossesP} Simulated photon flux in 7TeV centre-of-mass pp collisions at nominal luminosity shown for a quadrant of the CMS~\cite{CMS} experiment at the LHC~\cite{LHC}, CERN. The simulation uses the FLUKA code~\cite{FLUKA}. The coordinate system is shown relative to the interaction point for the pp collisions.  From~\cite{SMthesis}.}
\end{minipage}\hspace{0.5pc}%
\begin{minipage}[t]{19pc}
\includegraphics[width=19pc]{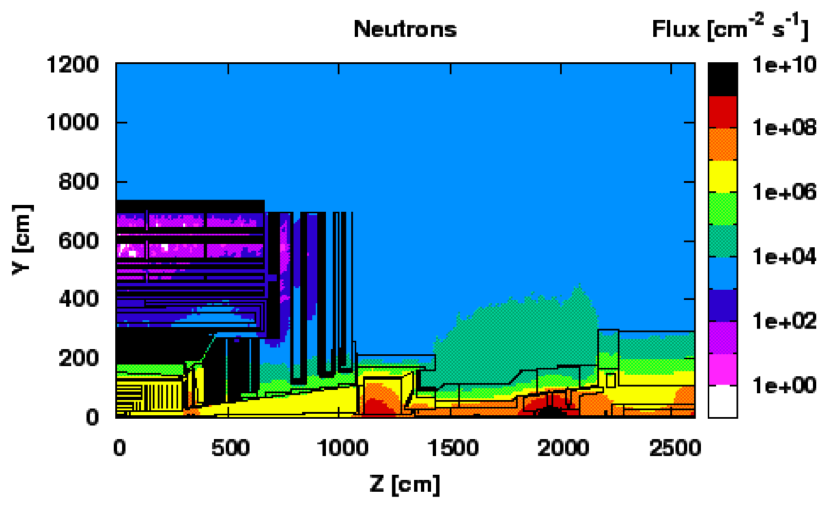}
\caption{\label{fig:CMSlossesN} Same as for figure~\ref{fig:CMSlossesP} but showing neutrons fluxes in the CMS experiment. From~\cite{SMthesis}}
\end{minipage} 
\end{figure}

When a straight beamline is necessary, it might even be optimal to allow the background to pass through unhindered.
Low pressure gaseous detectors, for example, can be quite insensitive to fast and epithermal neutrons and gammas~\cite{mgin6,b10gamma}; danger arises when material budgets increase, moderating the neutrons and increasing the interaction probability of gammas. 
This would require specific selection of materials to minimise particular materials, in particular hydrogen containing ones - and local shielding against thermal neutrons around the detectors. 
An analogy to this possible approach can also be found in typical high energy physics experiments, such as those at the LHC. Here the radiation source is point-like --- and the most sensitive detector volumes are very close to the source and on the exterior. 
To reduce both background levels and dose to sensitive electronics, the material budget is carefully engineered to produce a well-understood and manageable radiation flux in all detector components. 
An example from the CMS experiment is shown in figures~\ref{fig:CMSlossesP}~and~\ref{fig:CMSlossesN}, showing that sensitive components close in can have fluxes many orders of magnitude below those further out. 
A similar concept may help instruments requiring straight beamlines.

It is also the case that for the highest energy particles, there may be a shower. 
This may mean that a loss point close to, eg, an instrument cave, a too-thin collimator or a $T_0$ chopper could easily increase rather than decrease the background. 
It is therefore important to define well for what location the background is to be minimised, and design for that location the best background possible. 
Typically this means removing loss locations as far from this sensitive location as possible. 
This can be simply summarised as {\it ``choosing the loss point''} for undesired particles.


\section{Material}

As mentioned in section \ref{sec:shielding}, the traditional choice has been to converge on iron and concrete as shielding materials; typically this is seen as only a civil engineering task. Often this choice is dictated by the standard concrete block which exists at the facility. 
With a green-field site, a greater degree of flexibility exists, and should be investigated whether a less brute force approach can be applied. 

Additionally hydrogen-containing substances are very useful as they moderate fast neutrons. However, close to guides they may be detrimental, as moderated neutrons scattered back into the guide may give background effects with a long-tailed time structure, and they are also rather transparent to the high energy neutrons because of the low density. 
Care should be taken in their placement after the first set of choppers so that they are behind a barrier layer of rich boron material or other good thermal absorber, preventing moderated neutrons from escaping back out and further lengthening the tail of the pulse.  Another risk with hydrogenous materials that must be mitigated is the hazard of fire.  It is important at a green field site to consider plastics and waxes early so that the building systems can cope with the potential fire load.

\begin{figure}[h]
\begin{minipage}{17pc}
\includegraphics[width=17pc]{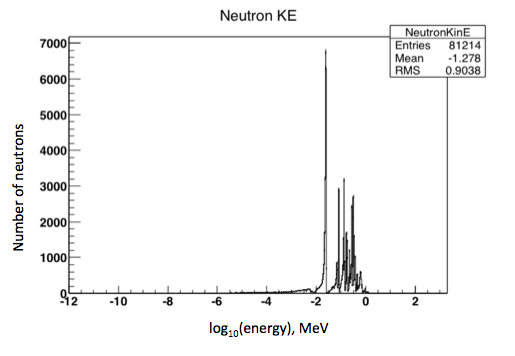}
\caption{\label{fig: iron} Energy spectrum of spallation neutrons attenuated by the 50 cm block of iron. Number of neutrons as a function of energy is shown.}
\end{minipage}\hspace{3pc}%
\begin{minipage}{17pc}
\includegraphics[width=17pc]{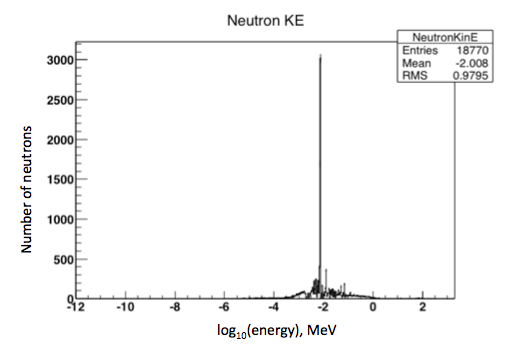}
\caption{\label{fig: copper}Energy spectrum of spallation neutrons attenuated by the 50 cm block of copper. Number of neutrons as a function of energy is shown.}
\end{minipage} 
\end{figure}

The ESS is now concentrating effort into the shielding properties of numerous different materials, such as specialised concrete, laminated materials, tungsten and copper-based solutions~\cite{copper}. The last ones include, for example, copper guides~\cite{copperguides}, or copper alloys such as brass. Copper
has been demonstrated at high energy physics labs in multiple roles.  It has superior shielding properties compared to iron and mild steels, due to the resonances in the total neutron cross section.
This is shown in figures \ref{fig: iron} and \ref{fig: copper}.
On the figure the simulated energy spectra of neutrons that came through the 50 cm blocks of those two materials are shown; spectrum used for the simulation was the ESS spallation spectrum.

Copper-based solutions deep inside key shielding areas are consequently seen to be attractive avenue of further research in reaching our background goals.
The aim is to use most appropriate and efficient dense material in the areas where moderating of the fast neutrons is essential for background suppression; shapes and volumes of shielding materials are also under investigation. A global cost-benefit study of materials and material combinations for key parts of the instrument shielding and guide bunker is well underway at ESS.

The integration of the neutron optics with the shielding strategy is crucial for a synergistic, high performance system.
Neutron guides are typically made of glass and surrounded at some distance by a vacuum tube: these are easy streaming paths for energetic neutrons as observed on CNCS at Spallation Neutron Source (ORNL).
Therefore, using copper close to guide, or as guide substrate to ensure that fast neutrons are slowed and stopped in as little material as possible is a very promising shielding option.  On the other hand, the glass guides are excellent escape or ``get lost'' points if they are purposely designed with the right bulk shielding geometry behind them, and interspersed periodically with denser materials. 

\begin{figure}[h]
\begin{minipage}{15pc}
\includegraphics[width=15pc]{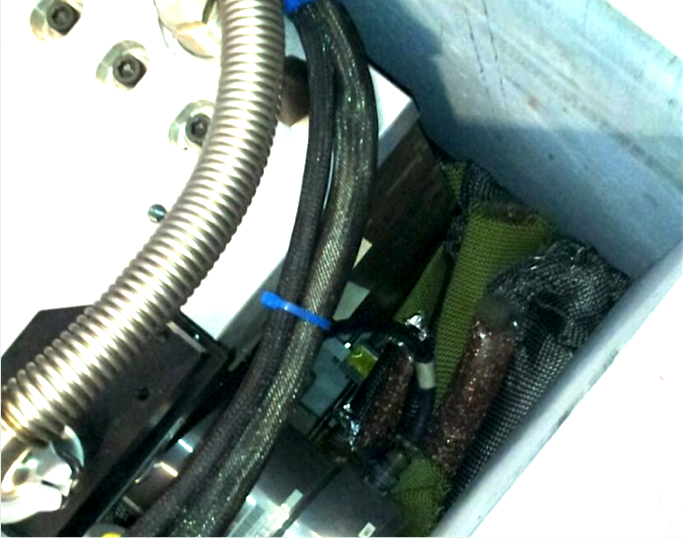}
\caption{\label{fig: cncs_photo} Bags with copper as an additional shielding at CNCS}
\end{minipage}\hspace{2pc}%
\begin{minipage}{19pc}
\includegraphics[width=19pc]{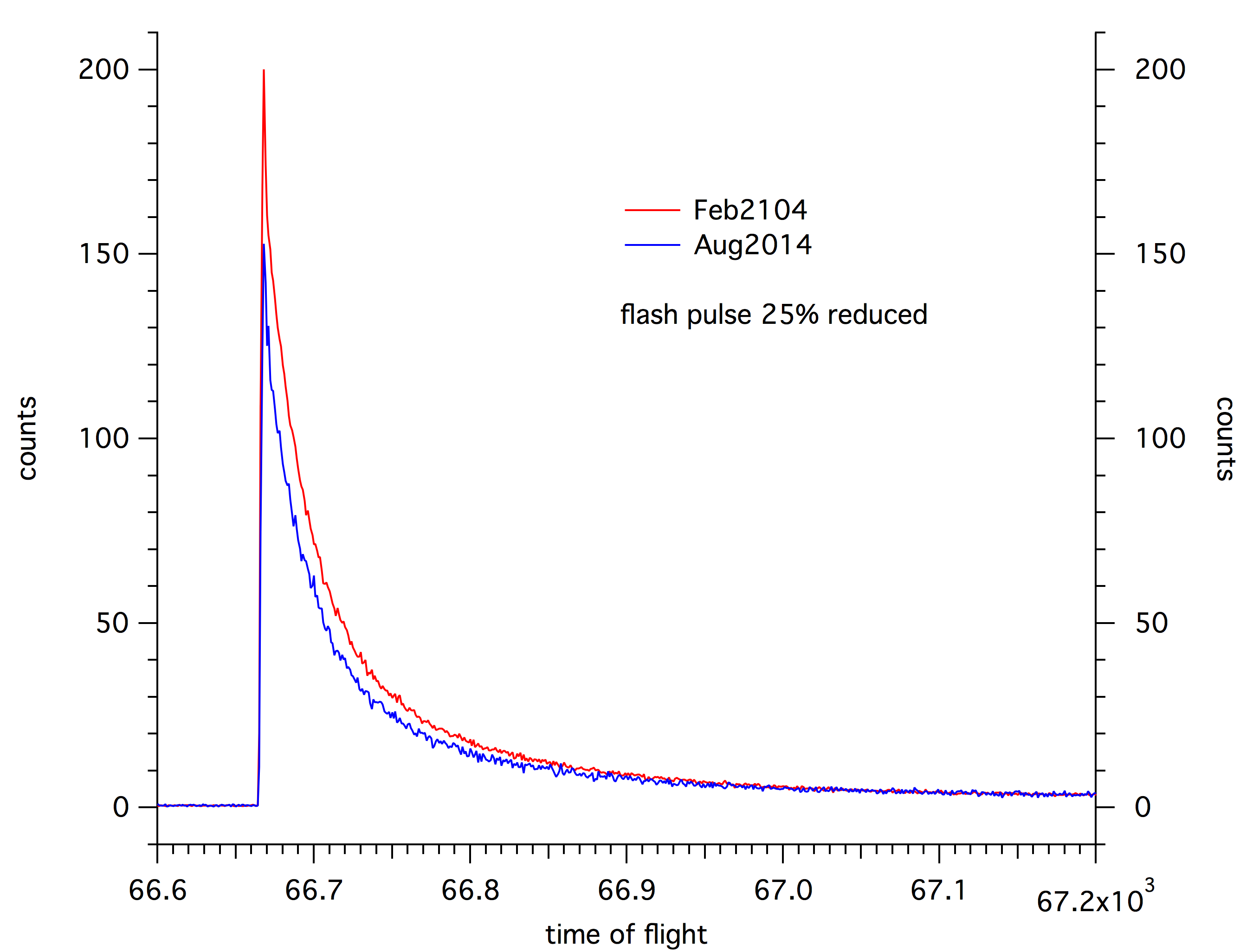}
\caption{\label{fig: cncs}Background reduction on CNCS achieved by adding copper in a few key places on the beamline.}
\end{minipage} 
\end{figure}

Recently copper was used in bags as a trial {\it ``bricolaged''} shielding improvement at CNCS (shown in figure \ref{fig: cncs_photo}),  and this brought a reduction in the prompt pulse background by 25$\%$, shown in figure \ref{fig: cncs}.

These early tests are very promising, and further reductions are anticipated with more work.  This highlights the great progress that the SNS team members are making in addressing instrument backgrounds of this kind, which --- despite being highly complex issues --- do not necessarily have to involve very expensive solutions if the choice of materials and their locations can be identified.  It should also be contrasted with the effort needed to increase the efficiency of modern, optimised guide sytstems by 25\%.

One more topic to discuss is the activation. Activation is an important issue in the choice of material, as neutrons activate material easily.
This is important to reduce decommissioning costs, and allow access to locations for maintenance.
Here both the desired material and its impurities should be considered. Steel is in principle good in that aspect, however there are many cases where unforeseeable impurities have created extreme activation problems on relatively moderate doses~\cite{gregoire,kojima}.
Copper has a reputation for activating greatly; however studies~\cite{huhtinen} for the CMS experiment at the LHC showed that over a longer time frame, as the daughter products were short lived, the activation can be lower than other materials - even lower than aluminium after several months for the CMS radiation field.
Indeed, whilst the shielding performance of copper at high energy is comparable to stainless steel, the activation of copper is far superior to stainless steel.
It is vital to verify this result with actual simulations for the expected ESS neutron spectra and radiation field.

Here, the non-functional requirements on manual access and maintenance targets are needed, in combination with extensive simulation of activation and lifetime, taking particular note of measured impurity levels to produce a known predictable result. 
Doing so avoids much heartache later.

\section{Summary}

We have presented a summary of the efforts currently underway in a collaboration between several laboratories, to address issues in neutron instrument performance.
What remains a priority for the immediate future is a combined, holistic approach involving all of these facets. 

Working on the shielding will have a significant effect on the instrument performance; and is a larger potential source of gain in instrument performance nowadays than optimised optics, which is frequently around --- or beyond --- 80\% designed brilliance transfer for modern concepts and technology.
Further brilliance transfer gains will become increasingly expensive with smaller returns in performance, however by more advanced shielding, it is possible to both reduce costs and to reduce backgrounds.
Reduced backgrounds directly improves instrument performance, as the figure of merit for most instrument design is most closely correlated to functions of $S/B$.
For many instrument categories, it is possible to imagine an order of magnitude improvement to existing designs.  This is in fact a design goal for the instruments at the ESS.

At first glance, such a strategy might seem to indicate increased total costs, but this is not the case.  An effective system design, placing the materials in the right places, reduces the total volume of material required and actually reduces the total system cost.  This allows greater investment in other parts of the instrument, such as custom sample environment, and more expensive and reliable mechanical devices that translate into less maintenance and downtime to repair faults.  These enhance instrument productivity and performance in slightly less tangible or less directly-visible ways as raw neutron flux gains have in the past, but are significant in the long term and should not be underestimated.

We have described how the pulsed background is comprised of a complex admixture of fast, epithermal and thermal neutrons and gammas.  Therefore, to achieve the desired performance gains, a holistic, horizontal and combined approach to the shielding appears to be most sensible, cutting across both optics and shielding for a new source such as ESS --- and is in fact the strategy for the Neutron Optics and Shielding group~\cite{nosg}, taking inspiration from the Paul Scherrer Institute in Switzerland.
But more than this, closely linking the work with the Detector design and Chopper interface management is absolutely essential.  
It also cuts across the design of the instrument suite as a whole, and the way that the integration projects are run at the facility level.
It is not just about ``radiation-protection'' in fact it is far beyond it; though radiation safety and personal protection are paramount, typically, the levels of sensitivity of the instrument are at least an order of magnitude beyond these requirements.
This is because the peak, pulsed background levels are so high compared to the time-averaged background levels that, by achieving prompt-pulse background objectives at pulsed spallation sources, the safety objectives should be already exceeded.

All of these aspects take time, patience, thought and above all validation in depth.

Significant synergy now exists between the spallation sources on this topic, and it is to the benefit of the entire community to exploit this collaboration.

\section*{References}
\bibliography{reference}

\end{document}